\documentclass[%
reprint,
superscriptaddress,
nofootinbib,
 amsmath,amssymb,
aps,
pre,
]{revtex4-2}

\usepackage{graphicx}
\usepackage{epstopdf}
\usepackage{dcolumn}
\usepackage{bm}

\begin{document}

\title{Magnetic anisotropy and critical behavior of the quaternary van der Waals ferromagnetic material $\bm{{\rm CrGe_{\delta}Si_{1-\delta}Te_3}}$}

\author{Zefang Li}
\affiliation{Beijing National Laboratory for Condensed Matter Physics, Institute of Physics, Chinese Academy of Sciences, Beijing 100190, China}
\affiliation{University of Chinese Academy of Sciences, Beijing 100049, China}

\author{Xue Li}
\affiliation{Beijing National Laboratory for Condensed Matter Physics, Institute of Physics, Chinese Academy of Sciences, Beijing 100190, China}
\affiliation{University of Chinese Academy of Sciences, Beijing 100049, China}

\author{Bei Ding}
\affiliation{Beijing National Laboratory for Condensed Matter Physics, Institute of Physics, Chinese Academy of Sciences, Beijing 100190, China}
\affiliation{University of Chinese Academy of Sciences, Beijing 100049, China}

\author{Hang Li}
\affiliation{Beijing National Laboratory for Condensed Matter Physics, Institute of Physics, Chinese Academy of Sciences, Beijing 100190, China}
\affiliation{University of Chinese Academy of Sciences, Beijing 100049, China}

\author{Yuan Yao}
\affiliation{Beijing National Laboratory for Condensed Matter Physics, Institute of Physics, Chinese Academy of Sciences, Beijing 100190, China}

\author{Xuekui Xi}\email{xi@iphy.ac.cn}
\affiliation{Beijing National Laboratory for Condensed Matter Physics, Institute of Physics, Chinese Academy of Sciences, Beijing 100190, China}

\author{Wenhong Wang}\email{wenhong.wang@iphy.ac.cn}
\affiliation{Beijing National Laboratory for Condensed Matter Physics, Institute of Physics, Chinese Academy of Sciences, Beijing 100190, China}
\affiliation{Songshan Lake Materials Laboratory, Dongguan, Guangdong 523808, China}

\date{\today}
\begin{abstract}
    Recently, two-dimensional ferromagnetism in the family of Chromium compounds $\rm CrXTe_3 (X=Si, Ge)$ has attracted a broad research interest. Despite the structural similarity in $\rm CrTe_6$ octahedra, the size effect of inserted Ge or Si dimer contributes to significant differences in magnetism. Here, we report a new quaternary van der Waals ferromagnetic material $\rm CrGe_{\delta}Si_{1-\delta}Te_3$ synthesized by flux method. Ge substitution in Si site results in the lattice expansion, further increasing the Curie temperature and reducing the magnetic anisotropy. The critical behavior of $\rm Cr_{0.96}Ge_{0.17}Si_{0.82}Te_3$ has been studied by specific heat as well as magnetization measurements. And the extracted critical exponents are self-consistent and well-obeying the scaling laws, which are closer to the 2D Ising model with interaction decaying as $J(r)\approx r^{-3.44}$.
\end{abstract}

\maketitle

\section{Introduction}

The recent discoveries of intrinsic magnetism in two-dimensional (2D) van der Waals (vdW) materials has promoted the development of technological and theoretical advances, such as the application of atomically thin and flexible magnetoelectric devices, the study of topology and phase transitions within quantum confinement, and many other novel physical phenomena \cite{SciRev,NatRev}. Although there are many high-throughput first-principles predictions for 2D magnetic materials \cite{DFTp1,DFTp2,DFTp3}, only few of them are experimentally confirmed. Therefore, experimental search for new 2D magnetic materials are especially important.

The family of Chromium compounds $\rm CrXTe_3 (X=Si,Ge)$ exhibits ferromagnetic semiconductor behavior. The electronic structure of $\rm CrGeTe_3$ (CGT) is directly measured by angle-resolved photoemission spectroscopy (ARPES) and indicate an indirect band gap of $\rm 0.38 \ eV$ \cite{ARPES1}. As for $\rm CrSiTe_3$ (CST), an optical property measurements give an indirect band gap of $\rm 0.4 \ eV$ and a direct gap of $\rm 1.2 \ eV$ \cite{Gap1,Gap2}. And the ARPES experiments of CST further confirm the Mott transition under the interplay of electronic correlations and magnetic ordering \cite{ARPES2}. The ferromagnetism of bulk CGT and CST originates from the Cr-Te-Cr superexchange interactions. The magnetic $\rm Cr^{3+}$ ions locate in a distorted octahedral crystal field with nearly quenched orbital angular momentum ($L \approx 0$), which results a spin-only magnetic moment $3.87 \ \rm \mu_B/Cr$ and a g-factor near 2. However, such weak spin-orbit coupling ($\xi \bm{L}\cdot\bm{S}$) from $\rm Cr^{3+}$ ionic state is not sufficient for maintaining a reasonable single-ion anisotropy to explain the giant magnetic anisotropy [$\rm K_{u}(5\ K)=4.78\times 10^4 \ J/m^3$ for CGT and $\rm K_{u}(5\ K)=11.49\times 10^4 \ J/m^3$ for CST] \cite{FMRli}. Therefore, the additional magnetic exchange anisotropy should be contributed by the ligand Te $5p$ spin-orbit coupling through the superexchange mechanism, and thus result in an anisotropic XXZ Heisenberg Hamiltonian \cite{ani}. The inelastic neutron scattering \cite{neuCST1,neuCST2} and critical exponent analysis study \cite{cbCGT1,cbCGT2,cbCGT3,cbCST} confirm the 2D ferromagnetic correlations even in bulk form. And our previous ferromagnetic resonance experiments of CGT and CST further rule out the isotropy Heisenberg Hamiltonian with small single-ion anistropy, since the isotropy part has no contribution to the observed critical-fluctuation-driven-g-factor anisotropy \cite{FMRli}.

Despite the similarity of local $\rm CrTe_6$ octahedral structure and  Cr-Te-Cr superexchange hopping pathways, the differences of magnetism mainly come from the Ge or Si sites that covalently bonded with Te and pushing the Cr-Te-Cr bond angle. The reduced magnetic anisotropy in CGT is attributed to additional Cr $3d$-unoccupied Te $5p$-Cr $3d$ superexchange channels opened by the smaller charge transfer energy \cite{ani}. The increasing Curie temperature $T_C$ from CST ($34 \ \rm K$) to CGT ($68 \ \rm K$) comes from two parts: On the one hand, the increase of the nearest-neighbor Cr-Cr distance will reduce the antiferromagnetic-coupled Cr-Cr direct exchange \cite{DFTm1,neuCST1}; On the other hand, as expected in the Goodenough-Kanamori rule, pushing the Cr-Te-Cr bond angle closer to $90^\circ$ will increase the effect of the ferromagnetic superexchange \cite{neuCST1,supex}.

Therefore, the above size effect of Ge and Si prompts us to grow the chemically substituted $\rm CrGe_{\delta}Si_{1-\delta}Te_3$ (CGST), which has not been reported in the past. We examined the crystal structure using high-resolution transmission electron microscope (HRTEM) and single crystal X-ray diffraction. In order to understand the magnetic behavior, we performed specific heat and extensive magnetization measurements to investigating the critical behavior. We find the Ge substitution will increase the in-plane Cr-Cr distance, thus result in a higher $T_C$ ($35.06\ \rm K$) and reduce the magnetic anisotropy ($\rm K_{u}(5\ K)=10.28\times 10^4 \ J/m^3$ for CGST). And the critical exponents extracted from the high field region (for example, modified Arrott plot, Kouvel-Fisher plot and critical isotherm analysis) are in good agreement with 2D Ising model and indicate a long-range magnetic coupling. However, the Curie-Weiss fit with a small external magnetic field, as well as the isothermal magnetization for in-plane direction, both indicate a new magnetic phase in the low field region, which is probably ferrimagnetic coupled up to $300\ \rm K$.

\section{EXPERIMENT METHOD}
Bulk $\rm CrGe_{\delta}Si_{1-\delta}Te_3$ single crystals were prepared by the self-flux method with a molar ratio of Cr : Si : Ge : Te = 10 : x : (16-x) : 70. The mixture of chromium pieces (99.95 \% purity, Kurt J. Lesker), silicon pieces (99.999 \% purity, Kurt J. Lesker), germanium pieces (99.999 \% purity, Kurt J. Lesker), antimony ingot (99.99 \% purity, Alfa Aesar) were mounted in an alumina crucible, which was sealed inside a quartz tube under high vacuum ($10^{-4}\ \rm Pa$). The tube was placed inside a shaft furnace to react over a period of 12 $\rm h$ at 1150 $\rm ^\circ C$, and followed by cooling down to 700 $\rm ^\circ C$ with a rate of 4 $\rm ^\circ C/h$. At this temperature, excessive molten flux was centrifuged quickly. The plate-like crystals were shiny and soft, which could be easily exfoliated. Among these as-grown single crystals, the ratio of Ge was difficult to increase due to the occurrence of impure phases. Therefore, here we focus on the specific compositions of $\rm Cr_{0.96}Ge_{0.17}Si_{0.82}Te_3$ and $\rm Cr_{0.96}Ge_{0.20}Si_{0.70}Te_3$.

In order to ensure consistency, a hexagonal piece of crystal was chosen for all the following measurements. Energy dispersive X-ray spectroscopy (EDXS, equipped in Hitachi S-4800 microscope) was used to identify the chemical composition on nine different areas of the cleaved surfaces, which showed the averaged proportion of $\rm Cr_{0.96}Ge_{0.17}Si_{0.82}Te_3$. The atomic configuration was characterized by high-resolution transmission electron microscope (HRTEM, JEOL ARM200F).

The crystal structure of single crystal was solved from X-ray crystallographic analysis (Bruker D8 venture). The X-ray intensity data were measured ($\lambda = 0.71073 $ \AA) and  integrated with the Bruker SAINT software package using a narrow-frame algorithm. Data were corrected for absorption effects using the Multi-Scan method (SADABS). The structure was solved and refined using the Bruker SHELXTL Software Package, using the space group $R\bar{3}$ (148) with $\rm Z = 3$ for the formula unit $\rm Cr_2Ge_{0.26}Si_{1.74}Te_6$.

The heat capacity at zero field (PPMS-9T, Quantum Design physical properties measurement system), as well as the temperature and field dependent magnetization (MPMS, Quantum Design magnetic property measurement system) were charecterized. In consideration of the demagnetization effect, it should be noted that the external applied field has been corrected for the internal magnetic field as $H_{\rm int}=H_{\rm ext}-NM$, where $N$ is the demagnetization factor and $M$ is the measured magnetization.

\section{Results and Disscussion}
\subsection{Crystal Structure Characterization}
\begin{figure}[!htbp]
    \includegraphics{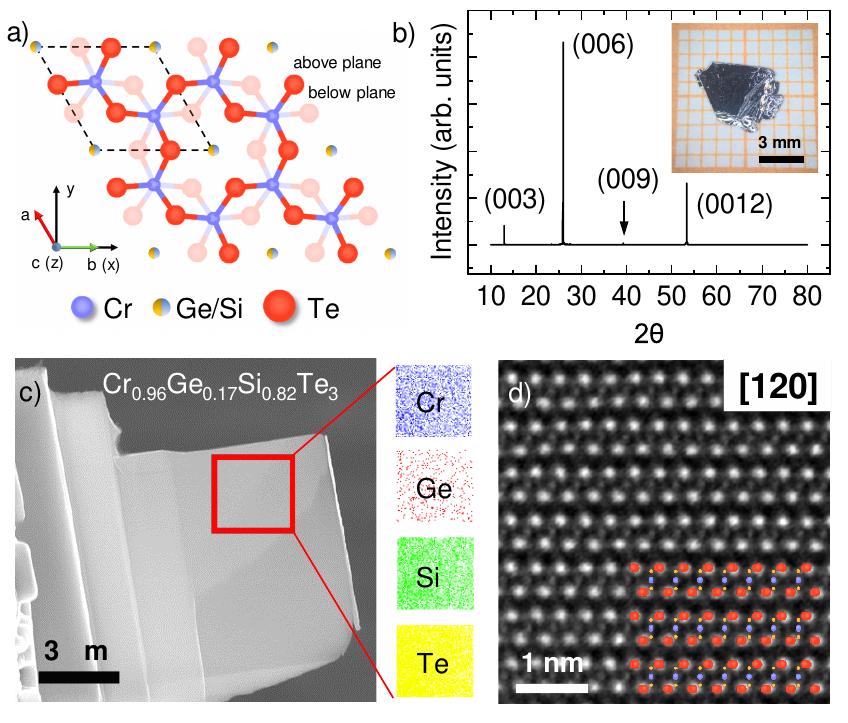}
    \caption{\label{fig:str}(a) Crystal structure of $\rm Cr_{0.96}Ge_{0.17}Si_{0.82}Te_3$ single crystal from the top view. (b) Single crystal x-ray diffraction and optical photograph of the single crystal. (c) SEM image and EDS mapping for the FIB prepared sample. (d) High-resolution STEM HAADF images along the [120] direction.}
\end{figure}

\begin{table}[!htbp]
    \caption{\label{tab:XRD1}%
    Sample and crystal data for $\rm Cr_{0.96}Ge_{0.17}Si_{0.82}Te_3$ at room temperature.}
    \begin{ruledtabular}
        \begin{tabular}{ccc}
            Refined formula & $\rm Cr_2Ge_{0.26}Si_{1.74}Te_6$ \\
            Formula weight & 937.14 $\rm g/mol$ \\
            Space group & $R\bar{3}$ (148) \\
            Unit cell dimensions & a = 6.7665(3) \AA & $\alpha=90^\circ$\\
                                 & b = 6.7665(3) \AA & $\beta=90^\circ$\\
                                 & c = 20.6281(15) \AA &$\gamma=90^\circ$\\
            Volume & 817.93(9) \AA$^3$ \\
            Density & 5.708 $\rm g/cm^3$
        \end{tabular}
        Atomic coordinates and equivalent isotropic atomic displacement parameters /\AA$^2$.
        \begin{tabular}{ccccc}
            & x & y & z & U(eq) \\
            Cr & 0.666667 & 0.333333 & 0.49924(5) & 0.01093(15) \\
            Ge & 0.0 & 0.0 & 0.55594(8) & 0.0107(4)\\
            Si & 0.0 & 0.0 & 0.55594(8) & 0.0107(4) \\
            Te & 0.36303(4) & 0.00246(4) & 0.58408(2) & 0.01161(8) \\
        \end{tabular}
    \end{ruledtabular}
\end{table}

Fig.\ref{fig:str}(a) shows the crystal structure of CGST single crystal, which is ABAB stacked by vdW gap. In each layer, the edge-shared $\rm CrTe_6$ octahedra forms a honeycomb network with Ge or Si dimer inserted. As shown in Tab. \ref{tab:XRD1}, the cell parameters of single crystal are defined by four-circle x-ray diffractometer. The space group is $R\bar{3}$, and lattice constants are $a=b=6.7665(3)$ \AA, $c = 20.6281(15)$ \AA. The sharp x-ray diffraction (XRD) patterns of ab plane in Fig. \ref{fig:str}(b) indicate high crystalline quality and no impurity phases. Fig. \ref{fig:str}(c) shows the foucused ion beam (FIB) prepared HRTEM samples for imaging along [120] direction. The EDXS mapping shows Ge is uniform doped in CST, with an averaged proportion of $\rm Cr_{0.96}Ge_{0.17}Si_{0.82}Te_3$ (slightly deviate from the XRD refined formula $\rm Cr_2Ge_{0.26}Si_{1.74}Te_6$). And the high-angle annular dark-field (HAADF) images in Fig. \ref{fig:str}(d) clearly shows the atomic arrangement is in good agreement with the refined structure. The above structural characterizations suggest that Ge substitution in CST is successful and the as-grown single crystal is near-perfect crystallized.

\subsection{Specific Heat}

\begin{figure}[!htbp]
    \includegraphics{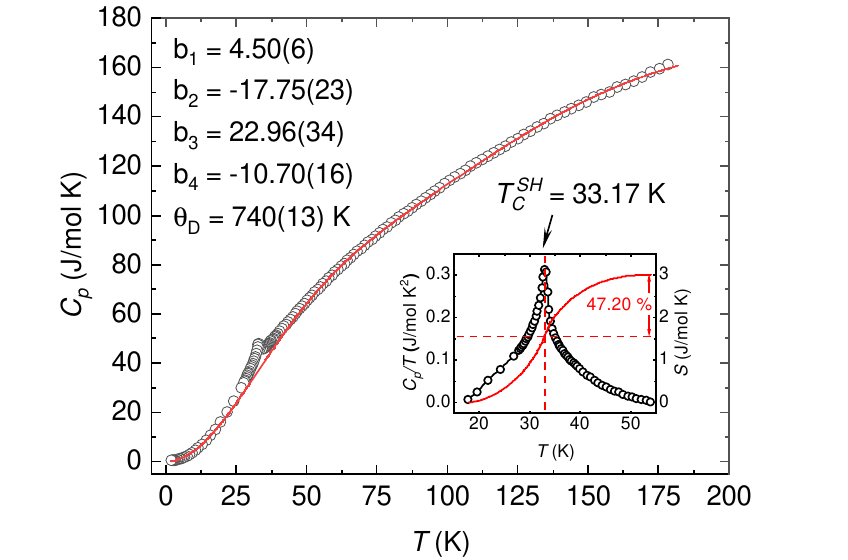}
    \caption{\label{fig:sh} Zero-field specific heat as a function of $T$ for $\rm Cr_{0.96}Ge_{0.17}Si_{0.82}Te_3$ and the fitted lattice contribution with Thirring model. The inset shows the magnetic contribution of $C_p/T$ versus $T$ and the integration of magnetic entropy $S$ after subtraction.}
\end{figure}

\begin{table}[!htbp]
    \caption{\label{tab:sh}%
    Comparison of zero-field specific heat data for different samples (See supplementary materials). \cite{FMRli}.}
    \begin{ruledtabular}
        \begin{tabular}{cccc}
            Chemical formula & $T_C^{SH}$ & magnetic entropy & above $T_C$\\ \hline
            $\rm CrSiTe_3$ & 32.68 $\rm K$ & 3.91 $\rm J/mol K$ & 48.67 \%\\
            $\rm Cr_{0.96}Ge_{0.17}Si_{0.82}Te_3$ & 33.17 $\rm K$ & 3.02 $\rm J/mol K$ & 47.20 \%\\
            $\rm Cr_{0.98}Ge_{0.20}Si_{0.70}Te_3$ & 36.68 $\rm K$ & 2.00 $\rm J/mol K$ & 38.17 \% \\
            $\rm CrGeTe_3$ & 64.90 $\rm K$ & 0.86 $\rm J/mol K$ & 38.85 \% \\
        \end{tabular}
    \end{ruledtabular}
\end{table}

Fig. \ref{fig:sh} shows the temperature dependence of zero-field specific heat. The sharp $\lambda$-shaped anomaly at $T_C^{SH} = 33.17\ {\rm K}$ corresponds to the PM-FM transition. The red solid line is the fitting of lattice contribution by Thirring model \cite{Thi}:
\begin{equation}
    \label{eq:Thirring}
    C_{\rm lattice}=3NR\left(1+\sum_{n=1}^\infty b_{n}\left(\left(\frac{2\pi T}{\theta_D}\right)^2+1\right)^{-n}\right),
\end{equation}
where $N$ is the number of atoms in the unit cell, $R$ is the ideal gas constant, $\theta_D$ is the Debye temperature. The series expansion $n=4$ is used for the fitting to obtain a reasonable accuracy. As shown in the inset, after subtracting the lattice contribution from the total heat capacity, we obtain the magnetic contribution and further integrating it to calculate the magnetic entropy change. Obviously, there is a large fraction of the magnetic entropies above $T_C$ ($47.20 \%$ for $\rm Cr_{0.96}Ge_{0.17}Si_{0.82}Te_3$). And the calculated magnetic entropy is relatively smaller than the theoretical value $S=k\ln W=R\ln(2J+1)=11.53\ {\rm J/mol \ K}$. This critical behavior near $T_c$ is attributed to the short-range magnetic correlations between the moments of nearest-neighbor atoms. As shown in Tab. \ref{tab:sh}, with the increase of Ge doping content, the $T_C^{SH}$ increases while magnetic entropy $S$ decreases, indicating the suppression of critical fluctuations near the transition temperature.

\subsection{M(T) and M(H) curves}

\begin{figure}[!htbp]
    \includegraphics{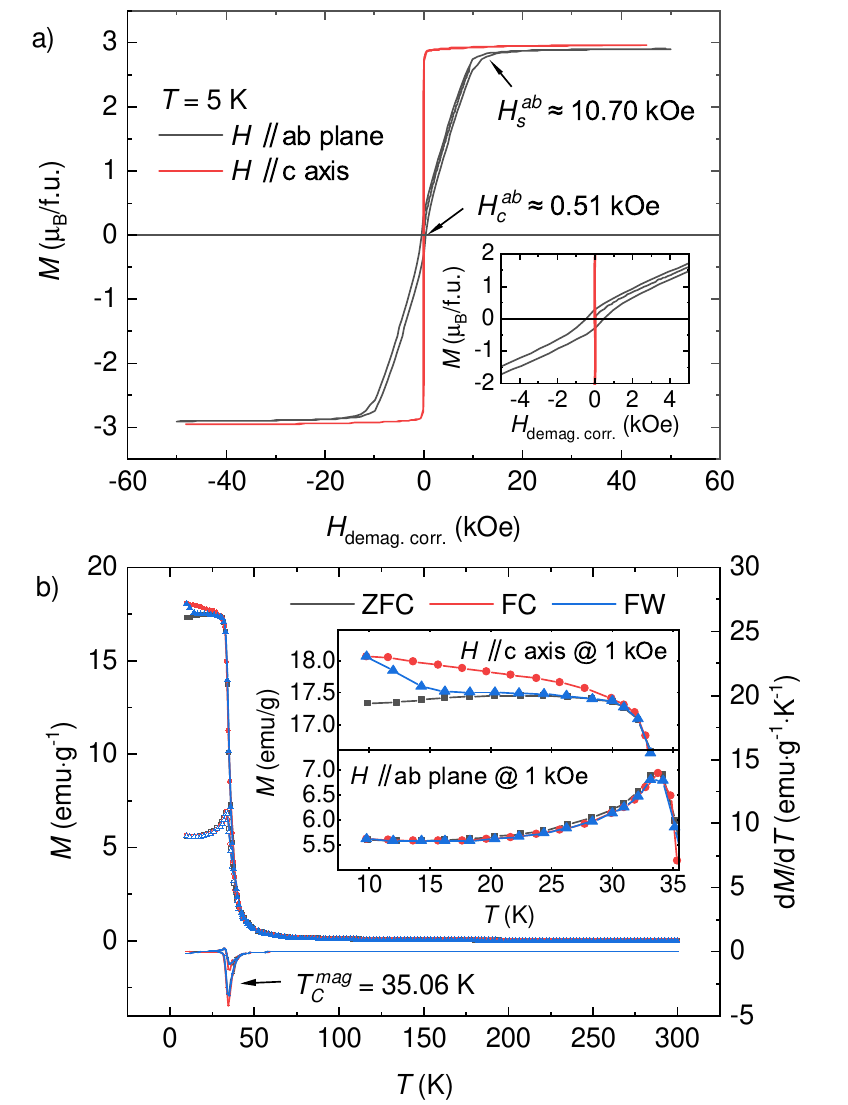}
    \caption{\label{fig:MHMT} (a) Field dependence of magnetization $M(H)$ measured at $T = 5 \ {\rm K}$. The demagnetization field is corrected according the sample shape. (b) Temperature dependence of the magnetization $M(T)$ and its derivative $dM/dT$ measured under zero field cooled (ZFC), field cooled (FC) and field warmming (FW) modes with a magnetic field $H = 1 \ {\rm kOe}$.}
\end{figure}

\begin{figure*}[!htbp]
    \includegraphics{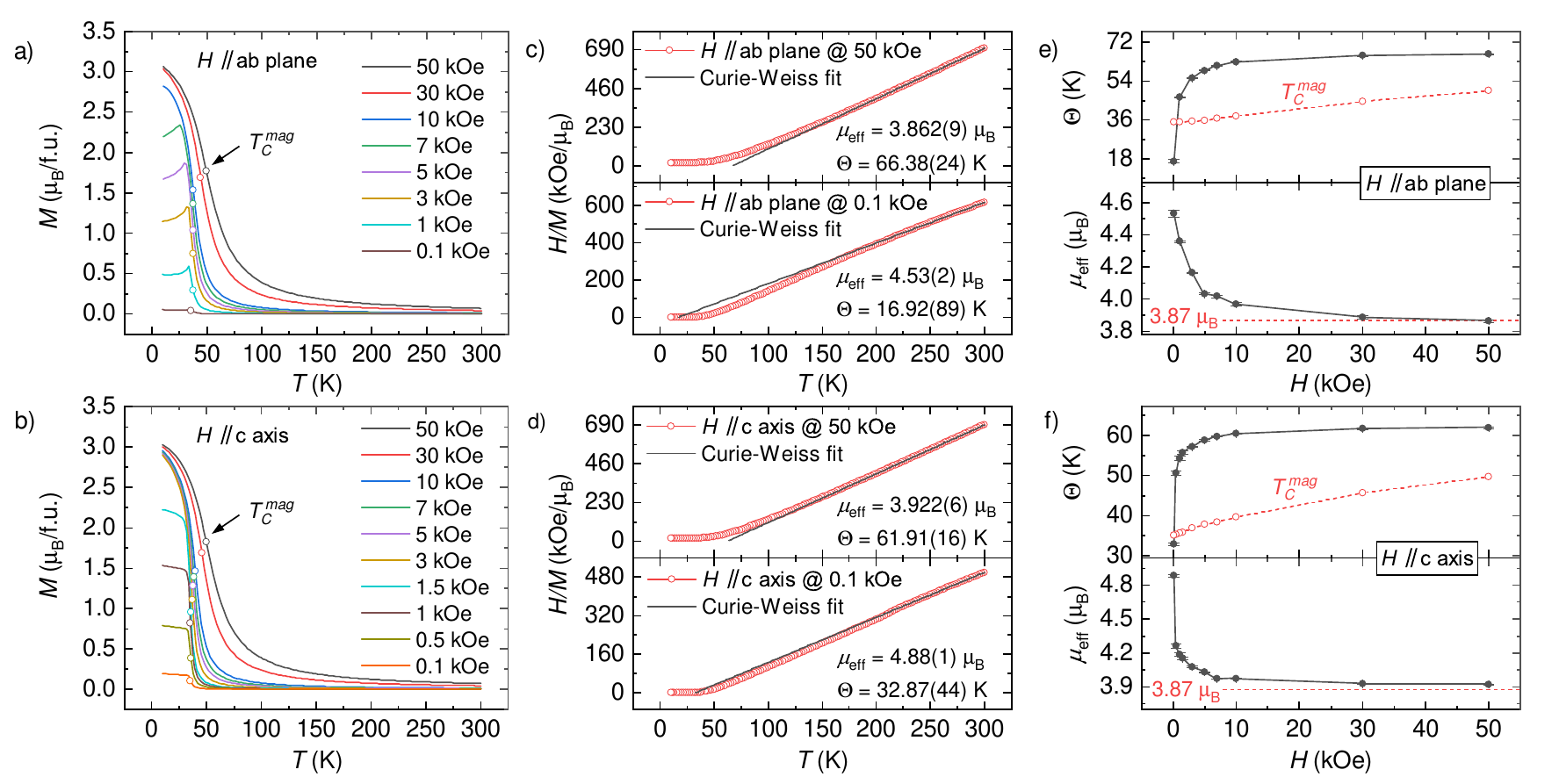}
    \caption{\label{fig:cw}(a), (b) Temperature dependence of in-plane ($H\parallel ab$ plane) and out-of-plane ($H\parallel c $ axis) magnetization under different magnetic fields. The hollow circles represents Curie temperature $T_c^{mag}$ determined by the minimum of $dM/dH$. (c), (d) The inverse susceptibility $H/M$ together with the Curie-Weiss fit under different magnetic fields. (e), (f) The fitted Curie-Weiss temperature $\Theta$ and effective magnetic moment $\mu_{\rm eff}$ under different magnetic fields.}
\end{figure*}

\begin{figure*}[!htbp]
    \includegraphics{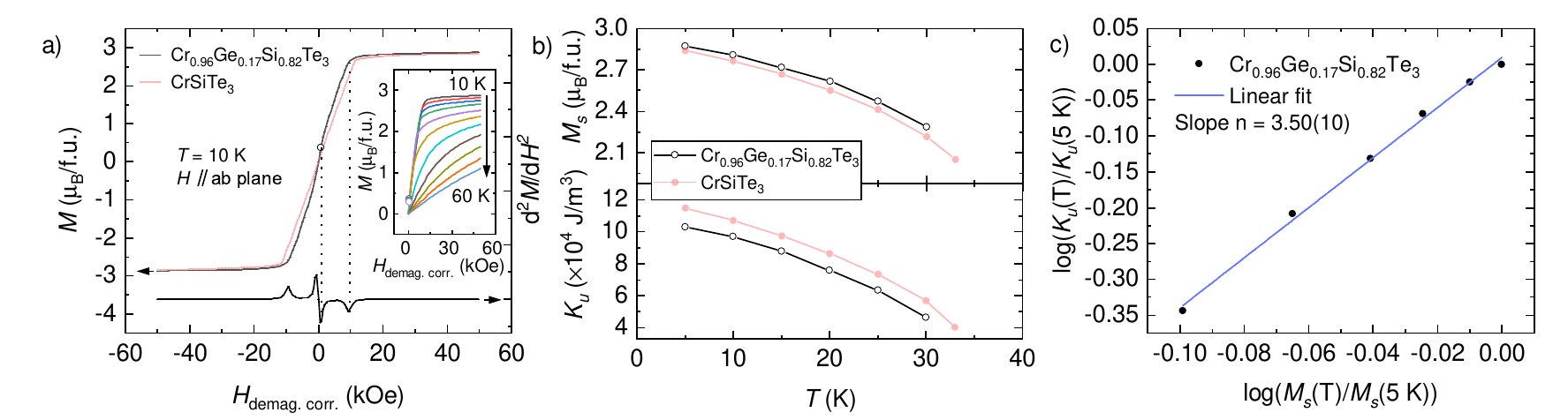}
    \caption{\label{fig:MH}(a) Comparison of the isothermal magnetization for $H \parallel ab$ plane at $5 \ \rm K$. The inset shows the isothermal magnetization of $\rm Cr_{0.96}Ge_{0.17}Si_{0.82}Te_3$ from $10 \ \rm K$ to $60 \ \rm K$. (b) Temperature dependence of saturated magnetization $M_s$ and anisotropy constant $K_u$. (c) lg-lg plot of the reduced anisotropy constant and magnetization with a linear fit for the Callen-Callen law.}
\end{figure*}

Fig. \ref{fig:MHMT}(a) shows the demagnetization-corrected isothermal magnetization $M(H)$ of $\rm Cr_{0.96}Ge_{0.17}Si_{0.82}Te_3$ at $5\ \rm K$ for both in-plane ($H \parallel ab$ plane) and out-of-plane ($H \parallel c$ axis) directions. The hysteresis of the out-of-plane magnetization is negligible, while there is a small coercive field of in-plane magnetization ($H_c^{ab} \approx 0.51 \ {\rm kOe}$). The $M(H)$ curves of CGST indicate a soft ferromagnet with an easy axis parallel to c axis. Fig. \ref{fig:MHMT}(b) shows the temperature dependence of magnetization $M(T)$ under an external magnetic field of $1\ \rm kOe$. The paramagnetic (PM) to ferromagnetic (FM) transition is determined by its derivative $dM/dT$. Compared with the Curie temperature of $\rm CrSiTe_3$ ($T_C^{mag} \approx 34.15 \ {\rm K}$) \cite{FMRli}, Ge doped CST shows a higher Curie temperature ($T_C^{mag} \approx 35.06 \ {\rm K}$ for $\rm Cr_{0.96}Ge_{0.17}Si_{0.82}Te_3$ and $T_C^{mag} \approx 39.15 \ {\rm K}$ for $\rm Cr_{0.98}Ge_{0.20}Si_{0.70}Te_3$). What's more, the real part of AC susceptibility $\chi'$ for $\rm Cr_{0.96}Ge_{0.17}Si_{0.82}Te_3$  shows a sharp anomaly at $34.13\ \rm K$, which is frequency independent (see supplementary materials).

The temperature dependent magnetization of FC is further characterized under different in-plane and out-of-plane magnetic fields [Fig. \ref{fig:cw}(a,b)]. The Curie-Weiss fits of the inverse susceptibility $H/M$ under the field of $50 \ \rm kOe$ and $0.1 \ \rm kOe$ are plotted in Fig. \ref{fig:cw}(c,d). As expected for $\rm Cr^{3+}$ ions in $\rm CrTe_6$ octahedra ($g=2$ and $J=S=3/2$), the theoretical value of effective magnetic moment should be $\mu_{\rm eff}=g\sqrt{J(J+1)}\mu_{\rm B}=3.87\ \mu_{\rm B}$. The fitted effective magnetic moment $\mu_{\rm eff}$ and Curie-Weiss temperature $\Theta$ are summarized in Fig. \ref{fig:cw}(e,f), which are gradually saturated in high field region. However, the obvious departures in low field region ($H < 10 \ {\rm kOe}$) indicate a new magnetic phase existing up to $300 \ \rm K$. Especially under the field of $0.1 \ \rm kOe$, the fitted Curie-Weiss temperature is below the $T_C^{mag}$. The most likely explanation is that Ge substitution could lead to ferrimagnetism at the doped site, but further verification is needed.

The anisotropic behavior of in-plane isothermal magnetization is compared with CST in Fig.\ref{fig:MH}(a). The initial magnetization of CST is almost a straight line, while a small bend in the low field region is observed in CGST. Such bend can be clearly seen from the second derivative $d^2M/dH^2$, and gradually disappears when near $T_C^{mag}$. The differences of CGST further bear out the new magnetic phase existing in the low field region. Moreover, Fig.\ref{fig:MH}(b) shows the saturation magnetization $M_s$ and magnetocrystalline anisotropy constant $K_u$ at different temperatures. From the high field region, the saturation magnetization at $5\ \rm K$ is in good agreement with the theoretical value $M_s=gJ\mu_{\rm B}=3.00\ \mu_{\rm B}/f.u.$. The anisotropy constant is calculated by the Stoner-Wolfarth model \cite{Ku}:
\begin{equation}
    \label{eq:ku}
    \frac{2K_u}{M_s}=\mu_0H_{\rm sat},
\end{equation}
where $\mu_0$ is the vacuum permeability, and $H_{\rm sat}$ is the saturation field. For the whole temperature range, we find the magnetic anisotropy is reduced after Ge substitution.

In addition, the Callen-Callen power law describes the ralation between the anisotropy and magnetization:
\begin{equation}
    \label{eq:CClaw}
    \frac{K_u(T)}{K_u(0)}=\left[\frac{M_s(T)}{M_s(0)}\right]^{n},
\end{equation}
where the exponent $n=l(l+1)/2$, and $l$ is the order of spherical harmonics and depends
on the symmetry of the crystal \cite{CClaw}, in the case of uniaxial anisotropy $n=3$ and of cubic anisotropy $n=10$. As shown in Fig. \ref{fig:MH}(c), we plot the reduced anisotropy constant and magnetization in the lg-lg scale. Thus a linear fit gives the exponent $n=3.50(10)$. The departure from Callen-Callen power law suggests the simple assumption of the single ion anisotropy is incomplete \cite{FMR1}. The additional magnetic exchange anisotropy should come elsewhere, such as the interplay between Kitaev interaction and single ion anisotropy \cite{Kitaev}, or anisotropic XXZ Heisenberg model \cite{FMRli,ani}.

\subsection{Critical Behavior}
\begin{figure}[!htbp]
    \includegraphics{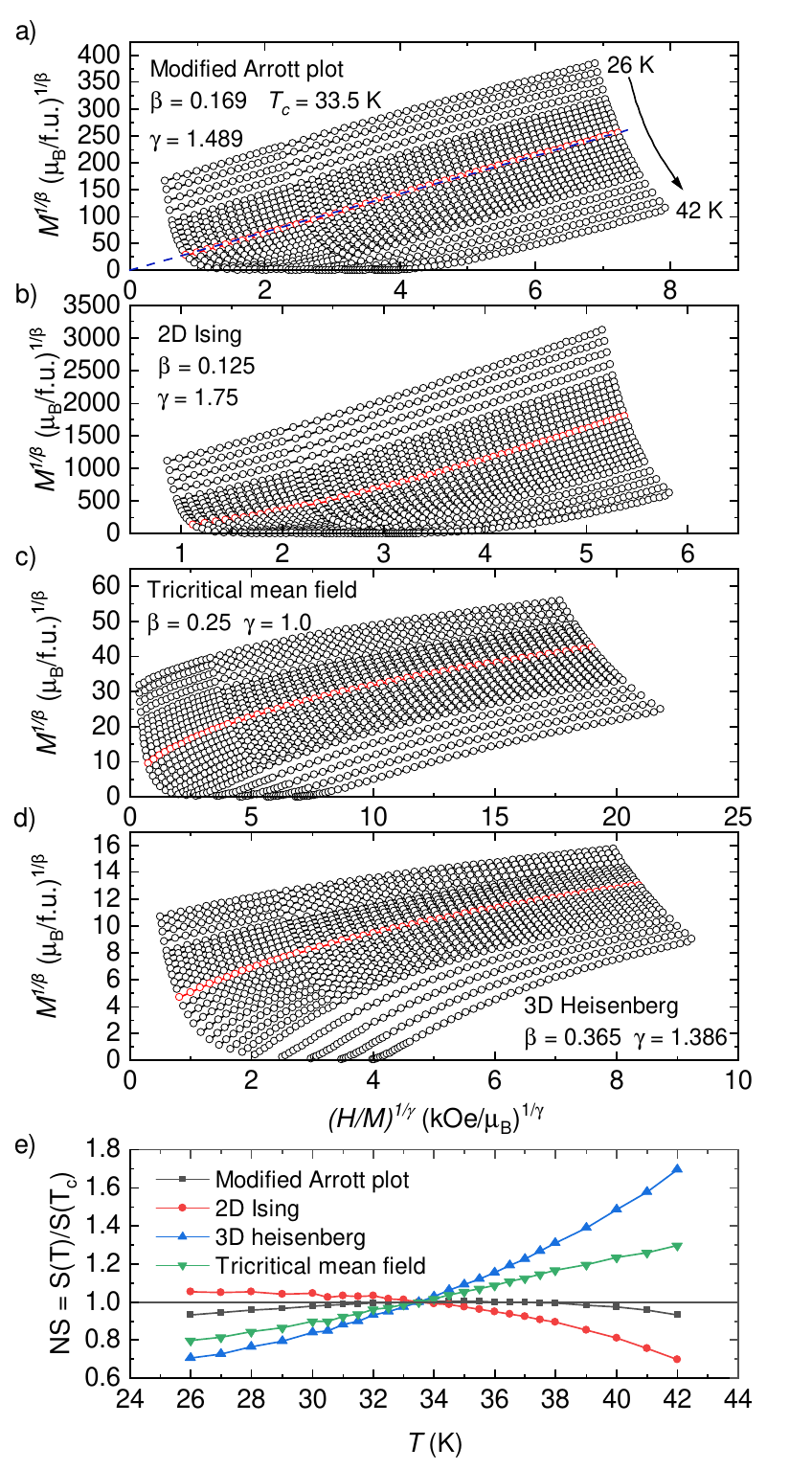}
    \caption{\label{fig:AP} The initial magnetization plotted as $M^{1/\beta}$ versus $(H/M)^{1/\gamma}$ based on (a) Modified Arrott plot, (b) 2D Ising model, (c) Tricritical mean-field model, and (d) 3D Heisenberg model. (e) Normalized slopes ${\rm NS} = S(T)/S(T_c)$ as a function of temperature.}
\end{figure}

\begin{figure*}[!htbp]
    \includegraphics{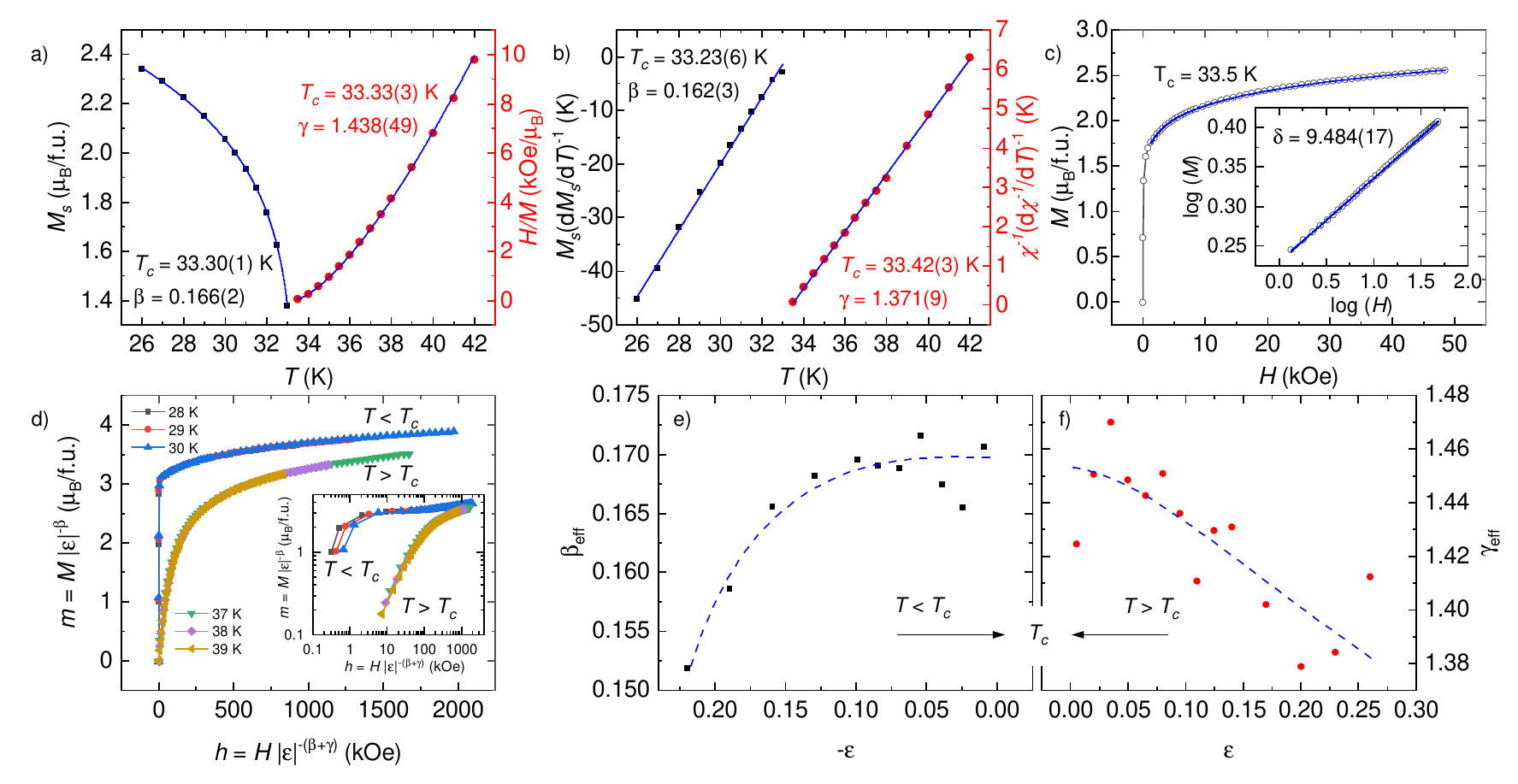}
    \caption{\label{fig:APa}(a) The temperature dependence of $M_s$ and $H/M$ with the solid line fitted by scaling law. (b) The Kouvel-Fisher plot of $M_s(dM_s/dT)^{-1}$ and $\chi^{-1}(d\chi^{-1}/dT)^{-1}$ with a linear fit. (c) Magnetic isothermal at $T_c$ with the solid line fitted by scaling law. The inset shows the lg-lg plot. (d) Scaling plots of renormalized magnetic isothermal around $T_c$. The inset shows the lg-lg plot for the same. (e), (f) Effective exponents of $\beta_{\rm eff}$ and $\gamma_{\rm eff}$ as a function of the reduced temperature $\varepsilon$.}
\end{figure*}

As for a magnetic system, the PM-FM transition is a second order phase transition and can be characterized by a series of interrelated critical exponents \cite{CE}. In the vicinity of the critical temperature, we define the reduced temperature $\varepsilon=(T-T_c)/T_c$. And the thermodynamic quantities near the critical point should follow the universal scaling laws, which can be mathematically defined as:
\begin{align}
    \label{eq:sclaw1}M_s(T) &\propto (-\varepsilon)^{\beta},  &(\varepsilon <0, T<T_c), \\
    \label{eq:sclaw2}\chi_0^{-1}(T) &\propto \varepsilon^{\gamma},  &(\varepsilon >0, T>T_c),\\
    \label{eq:sclaw3}M(H) & \propto H^{1/\delta}, &(\varepsilon =0, T=T_c),
\end{align}
where Equ. \ref{eq:sclaw1} is the spontaneous magnetization below $T_c$, Equ. \ref{eq:sclaw2} is the inverse susceptibility above $T_c$, Equ. \ref{eq:sclaw3} is critical isotherm at $T_c$. It should be noted that the critical exponents $\beta$, $\gamma$, $\delta$ are not independent:
\begin{equation}
    \label{eq:sclaw4}
    \delta=1+\frac{\gamma}{\beta}.
\end{equation}
In the vicinity of $T_c$, the experimental measured isothermal magnetization $M(H)$ are expected to collapse onto the scaling hypothesis, which be expressed as:
\begin{equation}
    \label{eq:sclaw5}
    M(H,\varepsilon)=\varepsilon^{\beta}f_{\pm}\left(\frac{H}{\varepsilon^{\beta+\gamma}}\right),
\end{equation}
where $f_{+}$ indicates the isotherms above $T_c$, and $f_{-}$ indicates the isotherms below $T_c$. And then we define renormalized magnetization $m\equiv \varepsilon^{-\beta}M(H,\varepsilon)$ and renormalized field $h\equiv \varepsilon^{-(\beta+\gamma)}H$. Equ. \ref{eq:sclaw5} can be simplified as:
\begin{equation}
    \label{eq:sclaw6}
    m=f_{\pm}(h),
\end{equation}
which suggests that the scaled $m(h)$ will fall onto two different universal curves (below and above $T_c$) if the critical exponents are chosen appropriately. In addition, it is useful to check the temperature dependent effective exponents for $\varepsilon \neq 0$:
\begin{equation}
    \label{eq:sclaw7}
    \beta_{\rm eff}(\varepsilon)=\frac{d[\ln M_s(\varepsilon)]}{d(\ln \varepsilon)}, \qquad \gamma_{\rm eff}(\varepsilon)=\frac{d[\ln \chi_0^{-1}(\varepsilon)]}{d(\ln \varepsilon)}
\end{equation}
As $\varepsilon \rightarrow 0$, which means the temperature approaches $T_c$, the effective exponents will approach universal exponents.

To further characterize the PM-FM transition and determine the critical exponents of CGST, we process the isothermal $M(H)$ with modified Arrott plot, which is given by the Arrott-Noakes equation of state \cite{AP}:
\begin{equation}
    \label{eq:mAP}
    \left(\frac{H}{M}\right)^{1/\gamma}=a\varepsilon+b\left(M\right)^{1/\beta},
\end{equation}
where $\varepsilon$ is the reduced temperature, $a$ and $b$ are constants. If the critical exponents $\gamma$ and $\beta$ are chosen appropriately, the modified Arrott plot will result in a series of parallel straight lines of $M^{1/\beta}$ versus $(H/M)^{1/\gamma}$, and the line at $T_c$ will pass through the origin. Fig. \ref{fig:AP}(a) shows the modified Arrott plot built with the values of $\beta=0.169$ and $\gamma=1.489$, which are obtained using the iteration method \cite{AP2}. It can be seen the plots are almost parallel straight lines. And the blue dashed line indicates the determined $T_c=33.5 \ {\rm K}$. In addition, 2D Ising model ($\beta=0.125$, $\gamma=1.75$), tricritical mean field model ($\beta=0.25$, $\gamma=1.0$), and 3D Heisenberg model ($\beta=0.365$, $\gamma=1.386$) \cite{APCST} are compared in Figs. \ref{fig:AP}(b-d). And the nonlinear curves suggest that the exponents for conventional model are not appropriate. To further compare the different models, we extract the slopes from the high field region and normalize as ${\rm NS}=S(T)/S(T_c)$. Fig. \ref{fig:AP}(e) shows the plots of $\rm NS$ versus $T$, revealing that our modified Arrott plot is close to the ideal value of 1 and is near the 2D Ising model.

We further verify the self-consistency of the critical exponents and $T_c$. As shown in Fig. \ref{fig:APa}(a), the linear extrapolations from high field region of the isotherms [see Fig. \ref{fig:AP}(a)] provide the intercepts on $M^{1/\beta'}$ and $(H/M)^{1/\gamma'}$. By fitting the intercepts to Equ. \ref{eq:sclaw1} and \ref{eq:sclaw2}, we can get the new values of $\beta=0.166(2)$ for $T_c=33.30(1) \ {\rm K}$ and $\gamma=1.438(49)$ for $T_c=33.33(3)\ {\rm K}$. Fig. \ref{fig:APa}(b) shows the critical exponents extracted by Kouvel-Fisher method \cite{KF}. By linear fitting of $M_s(dM_s/dT)^{-1}$ versus $T$ and $\chi_0^{-1}(d\chi_0^{-1}/dT)^{-1}$ versus $T$, we can obtain $1/\beta$ and $1/\gamma$ from the slopes, as well as $T_c$ from the intercepts. From the Kouvel-Fisher method the estimated exponents and $T_c$ are $\beta=0.162(3)$ for $T_c=33.23(6) \ {\rm K}$ and $\gamma=1.371(9)$ for $T_c=33.42(3)\ {\rm K}$, which match well with the results of modified Arrott plot. In addition, Fig. \ref{fig:AP}(c) shows critical isotherm $M(H)$ at $T_c$, and the inset presents the fitting of Equ. \ref{eq:sclaw3} in lg-lg scale. Thus we get $\delta=9.484(17)$, which is self consistent with the value of $\beta$ and $\gamma$ according the Equ. \ref{eq:sclaw4}.

In order to check whether the obtained critical exponents can generate the scaling equation of state by Equ. \ref{eq:sclaw5}, we plot the renormalized $m$ versus $h$ in Fig. \ref{fig:APa}(d). The isotherms collapse onto two universal curves, which indicate the reliability of our fitted critical exponents $\beta$ and $\gamma$. In addition, as shown in Fig. \ref{fig:APa}(e), the effective critical exponents has been calculated by Equ. \ref{eq:sclaw7}, which are convergent with the critical exponents when $\varepsilon$ approaches zero.

Finally, we would like to discuss the 2D nature of magnetic interactions in CGST. According the renormalization group theory analysis \cite{decay}, the interaction function $J(r)$ decays with distance $r$ as:
\begin{equation}
    \label{eq:decay}
    J(r)\approx r^{-(d+\sigma)},
\end{equation}
where $d$ is the spatial dimensionality and $\sigma$ is a positive constant indicating the range of the interaction in the system. $\sigma >2$ indicates a short range interaction, while $\sigma <2$ indicates a long range interaction. Moreover, the value of $\sigma$ can be estimated by the following equation:
\begin{equation}
    \label{eq:dim}
    \begin{split}
    \gamma=1+\frac{4(n+2)}{d(n+8)}\Delta\sigma+\frac{8(n+2)(n-4)}{d^2(n+8)^2} \\
    \times\left[ 1+\frac{2G(d/2)(7n+20)}{(n-4)(n+8)}\right]\Delta\sigma^2,
    \end{split}
\end{equation}
where $\Delta\sigma=(\sigma-d/2)$, $G(d/2)=3-\frac14(d/2)^2$, and $n$ is the spin dimensionality. As for the given $d:n$ values, putting the critical exponent $\gamma$ in Equ. \ref{eq:dim} will give $\sigma$. And the other exponents can be calculated from the following equations: $\nu=\gamma/\sigma$, $\alpha=2-\nu d$, $\beta=(2-\alpha-\gamma)/2$, $\delta=1+\gamma/\beta$. We repeat the process to find the best $d:n$ and $\sigma$ values that match the experimental critical exponents. For 3D Heisenberg-like spin system ($d=3$, $n=3$), Equ. \ref{eq:dim} gives $\sigma=2.02$ for $\gamma=1.48$, $\beta=0.36$.\, which suggests the spin interaction should not be 3D Heisenberg-like. However, 2D Ising type ($d=2$ $n=1$) spin interaction give the exponents ($\sigma=1.44$, $\gamma=1.49$ and $\beta=0.29$), which indicates a long range interaction. Therefore, the magnetic interaction in layered CGST behaves like 2D Ising type with the interaction decaying as $J(r)\approx r^{-3.44}$.

\section{CONCLUSION}
In summary, we report the new quaternary van der Waals ferromagnetic material $\rm CrGe_{\delta}Si_{1-\delta}Te_3$ grown by flux method for the first time. The structure and composition characterizations suggest that Ge substitution in CST is uniform, and the as-grown single crystal is near-perfect crystalized. Interestingly, the size effect of Ge substitution in Si site will increase the in-plane Cr-Cr distance, and thus increase the Curie temperature while decrease magnetic anisotropy. In the low field region, the anomaly of Curie-Weiss fit for $M(T)$, as well as the small bend in $M(H)$, probably indicate a new ferrimagnetic phase. The critical behavior of the PM-FM phase transition is comprehensive studied by modified Arrott plot, Kouvel-Fisher plot, and the critical isotherm analysis. The determined critical exponents ($\beta=0.169$,$\gamma=1.489$) are self-consistent and well-obeying the scaling laws. Furthermore, by comparing with the renormalization group calculations, the magnetic interaction of CGST behaves like 2D Ising tpye ($d=2$, $n=1$) with interaction decaying as $J(r)\approx r^{-3.44}$. The critical behavior of $\rm Cr_{0.96}Ge_{0.17}Si_{0.82}Te_3$ is close to that of CST \cite{cbCST}, but far away from CGT \cite{cbCGT1,cbCGT2,cbCGT3}. Moreover, according to Ginzburg criterion, such typical 2D magnetism is associated with strong intrinsic magnetization fluctuations when near the upper critical dimension \cite{Ginzburg}. The discovery of CGST offers a new platform for understanding the fluctuation-driven phase transition in CST, such as the pressure-induced superconductivity \cite{SC}, and the strain-induced Kitaev quantum spin liquid state \cite{QSL}.

\section*{ACKNOWLEDGMENTS}
This work was supported by the National Key R\&D Program of China (2017YFA0206303 and 2017YFA0303202), National Natural Science Foundation of China (Nrs. 51831003, 11974406, and 11874410), and Strategic Priority Research Program (B) of the Chinese Academy of Sciences (CAS) (XDB33000000).


\begin{thebibliography}{33}%
\makeatletter
\providecommand \@ifxundefined [1]{%
 \@ifx{#1\undefined}
}%
\providecommand \@ifnum [1]{%
 \ifnum #1\expandafter \@firstoftwo
 \else \expandafter \@secondoftwo
 \fi
}%
\providecommand \@ifx [1]{%
 \ifx #1\expandafter \@firstoftwo
 \else \expandafter \@secondoftwo
 \fi
}%
\providecommand \natexlab [1]{#1}%
\providecommand \enquote  [1]{``#1''}%
\providecommand \bibnamefont  [1]{#1}%
\providecommand \bibfnamefont [1]{#1}%
\providecommand \citenamefont [1]{#1}%
\providecommand \href@noop [0]{\@secondoftwo}%
\providecommand \href [0]{\begingroup \@sanitize@url \@href}%
\providecommand \@href[1]{\@@startlink{#1}\@@href}%
\providecommand \@@href[1]{\endgroup#1\@@endlink}%
\providecommand \@sanitize@url [0]{\catcode `\\12\catcode `\$12\catcode
  `\&12\catcode `\#12\catcode `\^12\catcode `\_12\catcode `\%12\relax}%
\providecommand \@@startlink[1]{}%
\providecommand \@@endlink[0]{}%
\providecommand \url  [0]{\begingroup\@sanitize@url \@url }%
\providecommand \@url [1]{\endgroup\@href {#1}{\urlprefix }}%
\providecommand \urlprefix  [0]{URL }%
\providecommand \Eprint [0]{\href }%
\providecommand \doibase [0]{https://doi.org/}%
\providecommand \selectlanguage [0]{\@gobble}%
\providecommand \bibinfo  [0]{\@secondoftwo}%
\providecommand \bibfield  [0]{\@secondoftwo}%
\providecommand \translation [1]{[#1]}%
\providecommand \BibitemOpen [0]{}%
\providecommand \bibitemStop [0]{}%
\providecommand \bibitemNoStop [0]{.\EOS\space}%
\providecommand \EOS [0]{\spacefactor3000\relax}%
\providecommand \BibitemShut  [1]{\csname bibitem#1\endcsname}%
\let\auto@bib@innerbib\@empty
\bibitem [{\citenamefont {Gong}\ and\ \citenamefont {Zhang}(2019)}]{SciRev}%
  \BibitemOpen
  \bibfield  {author} {\bibinfo {author} {\bibfnamefont {C.}~\bibnamefont
  {Gong}}\ and\ \bibinfo {author} {\bibfnamefont {X.}~\bibnamefont {Zhang}},\
  }\href@noop {} {\bibfield  {journal} {\bibinfo  {journal} {Science}\ }\textbf
  {\bibinfo {volume} {363}} (\bibinfo {year} {2019})}\BibitemShut {NoStop}%
\bibitem [{\citenamefont {Burch}\ \emph {et~al.}(2018)\citenamefont {Burch},
  \citenamefont {Mandrus},\ and\ \citenamefont {Park}}]{NatRev}%
  \BibitemOpen
  \bibfield  {author} {\bibinfo {author} {\bibfnamefont {K.~S.}\ \bibnamefont
  {Burch}}, \bibinfo {author} {\bibfnamefont {D.}~\bibnamefont {Mandrus}},\
  and\ \bibinfo {author} {\bibfnamefont {J.-G.}\ \bibnamefont {Park}},\
  }\href@noop {} {\bibfield  {journal} {\bibinfo  {journal} {Nature}\ }\textbf
  {\bibinfo {volume} {563}},\ \bibinfo {pages} {47} (\bibinfo {year}
  {2018})}\BibitemShut {NoStop}%
\bibitem [{\citenamefont {Liu}\ \emph {et~al.}(2018{\natexlab{a}})\citenamefont
  {Liu}, \citenamefont {Sun}, \citenamefont {Liu},\ and\ \citenamefont
  {Meng}}]{DFTp1}%
  \BibitemOpen
  \bibfield  {author} {\bibinfo {author} {\bibfnamefont {H.}~\bibnamefont
  {Liu}}, \bibinfo {author} {\bibfnamefont {J.-T.}\ \bibnamefont {Sun}},
  \bibinfo {author} {\bibfnamefont {M.}~\bibnamefont {Liu}},\ and\ \bibinfo
  {author} {\bibfnamefont {S.}~\bibnamefont {Meng}},\ }\href
  {https://doi.org/10.1021/acs.jpclett.8b02783} {\bibfield  {journal} {\bibinfo
   {journal} {The Journal of Physical Chemistry Letters}\ }\textbf {\bibinfo
  {volume} {9}},\ \bibinfo {pages} {6709} (\bibinfo {year}
  {2018}{\natexlab{a}})},\ \Eprint
  {https://arxiv.org/abs/https://doi.org/10.1021/acs.jpclett.8b02783}
  {https://doi.org/10.1021/acs.jpclett.8b02783} \BibitemShut {NoStop}%
\bibitem [{\citenamefont {Chittari}\ \emph {et~al.}(2020)\citenamefont
  {Chittari}, \citenamefont {Lee}, \citenamefont {Banerjee}, \citenamefont
  {MacDonald}, \citenamefont {Hwang},\ and\ \citenamefont {Jung}}]{DFTp2}%
  \BibitemOpen
  \bibfield  {author} {\bibinfo {author} {\bibfnamefont {B.~L.}\ \bibnamefont
  {Chittari}}, \bibinfo {author} {\bibfnamefont {D.}~\bibnamefont {Lee}},
  \bibinfo {author} {\bibfnamefont {N.}~\bibnamefont {Banerjee}}, \bibinfo
  {author} {\bibfnamefont {A.~H.}\ \bibnamefont {MacDonald}}, \bibinfo {author}
  {\bibfnamefont {E.}~\bibnamefont {Hwang}},\ and\ \bibinfo {author}
  {\bibfnamefont {J.}~\bibnamefont {Jung}},\ }\href
  {https://doi.org/10.1103/PhysRevB.101.085415} {\bibfield  {journal} {\bibinfo
   {journal} {Phys. Rev. B}\ }\textbf {\bibinfo {volume} {101}},\ \bibinfo
  {pages} {085415} (\bibinfo {year} {2020})}\BibitemShut {NoStop}%
\bibitem [{\citenamefont {Zhu}\ \emph {et~al.}(2018)\citenamefont {Zhu},
  \citenamefont {Kong}, \citenamefont {Rhone},\ and\ \citenamefont
  {Guo}}]{DFTp3}%
  \BibitemOpen
  \bibfield  {author} {\bibinfo {author} {\bibfnamefont {Y.}~\bibnamefont
  {Zhu}}, \bibinfo {author} {\bibfnamefont {X.}~\bibnamefont {Kong}}, \bibinfo
  {author} {\bibfnamefont {T.~D.}\ \bibnamefont {Rhone}},\ and\ \bibinfo
  {author} {\bibfnamefont {H.}~\bibnamefont {Guo}},\ }\href
  {https://doi.org/10.1103/PhysRevMaterials.2.081001} {\bibfield  {journal}
  {\bibinfo  {journal} {Phys. Rev. Materials}\ }\textbf {\bibinfo {volume}
  {2}},\ \bibinfo {pages} {081001} (\bibinfo {year} {2018})}\BibitemShut
  {NoStop}%
\bibitem [{\citenamefont {Li}\ \emph {et~al.}(2018)\citenamefont {Li},
  \citenamefont {Wang}, \citenamefont {Guo}, \citenamefont {Gu}, \citenamefont
  {Sun}, \citenamefont {He}, \citenamefont {Zhou}, \citenamefont {Gu},
  \citenamefont {Nie},\ and\ \citenamefont {Pan}}]{ARPES1}%
  \BibitemOpen
  \bibfield  {author} {\bibinfo {author} {\bibfnamefont {Y.~F.}\ \bibnamefont
  {Li}}, \bibinfo {author} {\bibfnamefont {W.}~\bibnamefont {Wang}}, \bibinfo
  {author} {\bibfnamefont {W.}~\bibnamefont {Guo}}, \bibinfo {author}
  {\bibfnamefont {C.~Y.}\ \bibnamefont {Gu}}, \bibinfo {author} {\bibfnamefont
  {H.~Y.}\ \bibnamefont {Sun}}, \bibinfo {author} {\bibfnamefont
  {L.}~\bibnamefont {He}}, \bibinfo {author} {\bibfnamefont {J.}~\bibnamefont
  {Zhou}}, \bibinfo {author} {\bibfnamefont {Z.~B.}\ \bibnamefont {Gu}},
  \bibinfo {author} {\bibfnamefont {Y.~F.}\ \bibnamefont {Nie}},\ and\ \bibinfo
  {author} {\bibfnamefont {X.~Q.}\ \bibnamefont {Pan}},\ }\href
  {https://doi.org/10.1103/PhysRevB.98.125127} {\bibfield  {journal} {\bibinfo
  {journal} {Phys. Rev. B}\ }\textbf {\bibinfo {volume} {98}},\ \bibinfo
  {pages} {125127} (\bibinfo {year} {2018})}\BibitemShut {NoStop}%
\bibitem [{\citenamefont {Casto}\ \emph {et~al.}(2015)\citenamefont {Casto},
  \citenamefont {Clune}, \citenamefont {Yokosuk}, \citenamefont {Musfeldt},
  \citenamefont {Williams}, \citenamefont {Zhuang}, \citenamefont {Lin},
  \citenamefont {Xiao}, \citenamefont {Hennig}, \citenamefont {Sales},
  \citenamefont {Yan},\ and\ \citenamefont {Mandrus}}]{Gap1}%
  \BibitemOpen
  \bibfield  {author} {\bibinfo {author} {\bibfnamefont {L.~D.}\ \bibnamefont
  {Casto}}, \bibinfo {author} {\bibfnamefont {A.~J.}\ \bibnamefont {Clune}},
  \bibinfo {author} {\bibfnamefont {M.~O.}\ \bibnamefont {Yokosuk}}, \bibinfo
  {author} {\bibfnamefont {J.~L.}\ \bibnamefont {Musfeldt}}, \bibinfo {author}
  {\bibfnamefont {T.~J.}\ \bibnamefont {Williams}}, \bibinfo {author}
  {\bibfnamefont {H.~L.}\ \bibnamefont {Zhuang}}, \bibinfo {author}
  {\bibfnamefont {M.-W.}\ \bibnamefont {Lin}}, \bibinfo {author} {\bibfnamefont
  {K.}~\bibnamefont {Xiao}}, \bibinfo {author} {\bibfnamefont {R.~G.}\
  \bibnamefont {Hennig}}, \bibinfo {author} {\bibfnamefont {B.~C.}\
  \bibnamefont {Sales}}, \bibinfo {author} {\bibfnamefont {J.-Q.}\ \bibnamefont
  {Yan}},\ and\ \bibinfo {author} {\bibfnamefont {D.}~\bibnamefont {Mandrus}},\
  }\href {https://doi.org/10.1063/1.4914134} {\bibfield  {journal} {\bibinfo
  {journal} {APL Materials}\ }\textbf {\bibinfo {volume} {3}},\ \bibinfo
  {pages} {041515} (\bibinfo {year} {2015})},\ \Eprint
  {https://arxiv.org/abs/https://doi.org/10.1063/1.4914134}
  {https://doi.org/10.1063/1.4914134} \BibitemShut {NoStop}%
\bibitem [{\citenamefont {Ji}\ \emph {et~al.}(2013)\citenamefont {Ji},
  \citenamefont {Stokes}, \citenamefont {Alegria}, \citenamefont {Blomberg},
  \citenamefont {Tanatar}, \citenamefont {Reijnders}, \citenamefont {Schoop},
  \citenamefont {Liang}, \citenamefont {Prozorov}, \citenamefont {Burch},
  \citenamefont {Ong}, \citenamefont {Petta},\ and\ \citenamefont
  {Cava}}]{Gap2}%
  \BibitemOpen
  \bibfield  {author} {\bibinfo {author} {\bibfnamefont {H.}~\bibnamefont
  {Ji}}, \bibinfo {author} {\bibfnamefont {R.~A.}\ \bibnamefont {Stokes}},
  \bibinfo {author} {\bibfnamefont {L.~D.}\ \bibnamefont {Alegria}}, \bibinfo
  {author} {\bibfnamefont {E.~C.}\ \bibnamefont {Blomberg}}, \bibinfo {author}
  {\bibfnamefont {M.~A.}\ \bibnamefont {Tanatar}}, \bibinfo {author}
  {\bibfnamefont {A.}~\bibnamefont {Reijnders}}, \bibinfo {author}
  {\bibfnamefont {L.~M.}\ \bibnamefont {Schoop}}, \bibinfo {author}
  {\bibfnamefont {T.}~\bibnamefont {Liang}}, \bibinfo {author} {\bibfnamefont
  {R.}~\bibnamefont {Prozorov}}, \bibinfo {author} {\bibfnamefont {K.~S.}\
  \bibnamefont {Burch}}, \bibinfo {author} {\bibfnamefont {N.~P.}\ \bibnamefont
  {Ong}}, \bibinfo {author} {\bibfnamefont {J.~R.}\ \bibnamefont {Petta}},\
  and\ \bibinfo {author} {\bibfnamefont {R.~J.}\ \bibnamefont {Cava}},\ }\href
  {https://doi.org/10.1063/1.4822092} {\bibfield  {journal} {\bibinfo
  {journal} {Journal of Applied Physics}\ }\textbf {\bibinfo {volume} {114}},\
  \bibinfo {pages} {114907} (\bibinfo {year} {2013})},\ \Eprint
  {https://arxiv.org/abs/https://doi.org/10.1063/1.4822092}
  {https://doi.org/10.1063/1.4822092} \BibitemShut {NoStop}%
\bibitem [{\citenamefont {Zhang}\ \emph {et~al.}(2019)\citenamefont {Zhang},
  \citenamefont {Cai}, \citenamefont {Xia}, \citenamefont {Liang},
  \citenamefont {Huang}, \citenamefont {Wang}, \citenamefont {Yang},
  \citenamefont {Yuan}, \citenamefont {Chen}, \citenamefont {Zhang},
  \citenamefont {Guo}, \citenamefont {Liu},\ and\ \citenamefont {Li}}]{ARPES2}%
  \BibitemOpen
  \bibfield  {author} {\bibinfo {author} {\bibfnamefont {J.}~\bibnamefont
  {Zhang}}, \bibinfo {author} {\bibfnamefont {X.}~\bibnamefont {Cai}}, \bibinfo
  {author} {\bibfnamefont {W.}~\bibnamefont {Xia}}, \bibinfo {author}
  {\bibfnamefont {A.}~\bibnamefont {Liang}}, \bibinfo {author} {\bibfnamefont
  {J.}~\bibnamefont {Huang}}, \bibinfo {author} {\bibfnamefont
  {C.}~\bibnamefont {Wang}}, \bibinfo {author} {\bibfnamefont {L.}~\bibnamefont
  {Yang}}, \bibinfo {author} {\bibfnamefont {H.}~\bibnamefont {Yuan}}, \bibinfo
  {author} {\bibfnamefont {Y.}~\bibnamefont {Chen}}, \bibinfo {author}
  {\bibfnamefont {S.}~\bibnamefont {Zhang}}, \bibinfo {author} {\bibfnamefont
  {Y.}~\bibnamefont {Guo}}, \bibinfo {author} {\bibfnamefont {Z.}~\bibnamefont
  {Liu}},\ and\ \bibinfo {author} {\bibfnamefont {G.}~\bibnamefont {Li}},\
  }\href {https://doi.org/10.1103/PhysRevLett.123.047203} {\bibfield  {journal}
  {\bibinfo  {journal} {Phys. Rev. Lett.}\ }\textbf {\bibinfo {volume} {123}},\
  \bibinfo {pages} {047203} (\bibinfo {year} {2019})}\BibitemShut {NoStop}%
\bibitem [{\citenamefont {Li}\ \emph {et~al.}(2021)\citenamefont {Li},
  \citenamefont {Xu}, \citenamefont {Li}, \citenamefont {Liao}, \citenamefont
  {Xi}, \citenamefont {Yu},\ and\ \citenamefont {Wang}}]{FMRli}%
  \BibitemOpen
  \bibfield  {author} {\bibinfo {author} {\bibfnamefont {Z.}~\bibnamefont
  {Li}}, \bibinfo {author} {\bibfnamefont {D.-H.}\ \bibnamefont {Xu}}, \bibinfo
  {author} {\bibfnamefont {X.}~\bibnamefont {Li}}, \bibinfo {author}
  {\bibfnamefont {H.-J.}\ \bibnamefont {Liao}}, \bibinfo {author}
  {\bibfnamefont {X.}~\bibnamefont {Xi}}, \bibinfo {author} {\bibfnamefont
  {Y.-C.}\ \bibnamefont {Yu}},\ and\ \bibinfo {author} {\bibfnamefont
  {W.}~\bibnamefont {Wang}},\ }\href@noop {} {\bibfield  {journal} {\bibinfo
  {journal} {arXiv preprint arXiv:2101.02440}\ } (\bibinfo {year}
  {2021})}\BibitemShut {NoStop}%
\bibitem [{\citenamefont {Kim}\ \emph {et~al.}(2019)\citenamefont {Kim},
  \citenamefont {Kim}, \citenamefont {Ko}, \citenamefont {Seo}, \citenamefont
  {Kim}, \citenamefont {Jang}, \citenamefont {Kim}, \citenamefont {Kim},
  \citenamefont {Cheong},\ and\ \citenamefont {Park}}]{ani}%
  \BibitemOpen
  \bibfield  {author} {\bibinfo {author} {\bibfnamefont {D.-H.}\ \bibnamefont
  {Kim}}, \bibinfo {author} {\bibfnamefont {K.}~\bibnamefont {Kim}}, \bibinfo
  {author} {\bibfnamefont {K.-T.}\ \bibnamefont {Ko}}, \bibinfo {author}
  {\bibfnamefont {J.}~\bibnamefont {Seo}}, \bibinfo {author} {\bibfnamefont
  {J.~S.}\ \bibnamefont {Kim}}, \bibinfo {author} {\bibfnamefont {T.-H.}\
  \bibnamefont {Jang}}, \bibinfo {author} {\bibfnamefont {Y.}~\bibnamefont
  {Kim}}, \bibinfo {author} {\bibfnamefont {J.-Y.}\ \bibnamefont {Kim}},
  \bibinfo {author} {\bibfnamefont {S.-W.}\ \bibnamefont {Cheong}},\ and\
  \bibinfo {author} {\bibfnamefont {J.-H.}\ \bibnamefont {Park}},\ }\href
  {https://doi.org/10.1103/PhysRevLett.122.207201} {\bibfield  {journal}
  {\bibinfo  {journal} {Phys. Rev. Lett.}\ }\textbf {\bibinfo {volume} {122}},\
  \bibinfo {pages} {207201} (\bibinfo {year} {2019})}\BibitemShut {NoStop}%
\bibitem [{\citenamefont {Williams}\ \emph {et~al.}(2015)\citenamefont
  {Williams}, \citenamefont {Aczel}, \citenamefont {Lumsden}, \citenamefont
  {Nagler}, \citenamefont {Stone}, \citenamefont {Yan},\ and\ \citenamefont
  {Mandrus}}]{neuCST1}%
  \BibitemOpen
  \bibfield  {author} {\bibinfo {author} {\bibfnamefont {T.~J.}\ \bibnamefont
  {Williams}}, \bibinfo {author} {\bibfnamefont {A.~A.}\ \bibnamefont {Aczel}},
  \bibinfo {author} {\bibfnamefont {M.~D.}\ \bibnamefont {Lumsden}}, \bibinfo
  {author} {\bibfnamefont {S.~E.}\ \bibnamefont {Nagler}}, \bibinfo {author}
  {\bibfnamefont {M.~B.}\ \bibnamefont {Stone}}, \bibinfo {author}
  {\bibfnamefont {J.-Q.}\ \bibnamefont {Yan}},\ and\ \bibinfo {author}
  {\bibfnamefont {D.}~\bibnamefont {Mandrus}},\ }\href
  {https://doi.org/10.1103/PhysRevB.92.144404} {\bibfield  {journal} {\bibinfo
  {journal} {Phys. Rev. B}\ }\textbf {\bibinfo {volume} {92}},\ \bibinfo
  {pages} {144404} (\bibinfo {year} {2015})}\BibitemShut {NoStop}%
\bibitem [{\citenamefont {Carteaux}\ \emph {et~al.}(1995)\citenamefont
  {Carteaux}, \citenamefont {Moussa},\ and\ \citenamefont
  {Spiesser}}]{neuCST2}%
  \BibitemOpen
  \bibfield  {author} {\bibinfo {author} {\bibfnamefont {V.}~\bibnamefont
  {Carteaux}}, \bibinfo {author} {\bibfnamefont {F.}~\bibnamefont {Moussa}},\
  and\ \bibinfo {author} {\bibfnamefont {M.}~\bibnamefont {Spiesser}},\
  }\href@noop {} {\bibfield  {journal} {\bibinfo  {journal} {EPL (Europhysics
  Letters)}\ }\textbf {\bibinfo {volume} {29}},\ \bibinfo {pages} {251}
  (\bibinfo {year} {1995})}\BibitemShut {NoStop}%
\bibitem [{\citenamefont {Liu}\ and\ \citenamefont {Petrovic}(2017)}]{cbCGT1}%
  \BibitemOpen
  \bibfield  {author} {\bibinfo {author} {\bibfnamefont {Y.}~\bibnamefont
  {Liu}}\ and\ \bibinfo {author} {\bibfnamefont {C.}~\bibnamefont {Petrovic}},\
  }\href {https://doi.org/10.1103/PhysRevB.96.054406} {\bibfield  {journal}
  {\bibinfo  {journal} {Phys. Rev. B}\ }\textbf {\bibinfo {volume} {96}},\
  \bibinfo {pages} {054406} (\bibinfo {year} {2017})}\BibitemShut {NoStop}%
\bibitem [{\citenamefont {Liu}\ \emph {et~al.}(2018{\natexlab{b}})\citenamefont
  {Liu}, \citenamefont {Dai}, \citenamefont {Yang}, \citenamefont {Fan},
  \citenamefont {Pi}, \citenamefont {Zhang},\ and\ \citenamefont
  {Zhang}}]{cbCGT2}%
  \BibitemOpen
  \bibfield  {author} {\bibinfo {author} {\bibfnamefont {W.}~\bibnamefont
  {Liu}}, \bibinfo {author} {\bibfnamefont {Y.}~\bibnamefont {Dai}}, \bibinfo
  {author} {\bibfnamefont {Y.-E.}\ \bibnamefont {Yang}}, \bibinfo {author}
  {\bibfnamefont {J.}~\bibnamefont {Fan}}, \bibinfo {author} {\bibfnamefont
  {L.}~\bibnamefont {Pi}}, \bibinfo {author} {\bibfnamefont {L.}~\bibnamefont
  {Zhang}},\ and\ \bibinfo {author} {\bibfnamefont {Y.}~\bibnamefont {Zhang}},\
  }\href {https://doi.org/10.1103/PhysRevB.98.214420} {\bibfield  {journal}
  {\bibinfo  {journal} {Phys. Rev. B}\ }\textbf {\bibinfo {volume} {98}},\
  \bibinfo {pages} {214420} (\bibinfo {year} {2018}{\natexlab{b}})}\BibitemShut
  {NoStop}%
\bibitem [{\citenamefont {Lin}\ \emph {et~al.}(2017)\citenamefont {Lin},
  \citenamefont {Zhuang}, \citenamefont {Luo}, \citenamefont {Liu},
  \citenamefont {Chen}, \citenamefont {Yan}, \citenamefont {Sun}, \citenamefont
  {Zhou}, \citenamefont {Lu}, \citenamefont {Tong}, \citenamefont {Sheng},
  \citenamefont {Qu}, \citenamefont {Song}, \citenamefont {Zhu},\ and\
  \citenamefont {Sun}}]{cbCGT3}%
  \BibitemOpen
  \bibfield  {author} {\bibinfo {author} {\bibfnamefont {G.~T.}\ \bibnamefont
  {Lin}}, \bibinfo {author} {\bibfnamefont {H.~L.}\ \bibnamefont {Zhuang}},
  \bibinfo {author} {\bibfnamefont {X.}~\bibnamefont {Luo}}, \bibinfo {author}
  {\bibfnamefont {B.~J.}\ \bibnamefont {Liu}}, \bibinfo {author} {\bibfnamefont
  {F.~C.}\ \bibnamefont {Chen}}, \bibinfo {author} {\bibfnamefont
  {J.}~\bibnamefont {Yan}}, \bibinfo {author} {\bibfnamefont {Y.}~\bibnamefont
  {Sun}}, \bibinfo {author} {\bibfnamefont {J.}~\bibnamefont {Zhou}}, \bibinfo
  {author} {\bibfnamefont {W.~J.}\ \bibnamefont {Lu}}, \bibinfo {author}
  {\bibfnamefont {P.}~\bibnamefont {Tong}}, \bibinfo {author} {\bibfnamefont
  {Z.~G.}\ \bibnamefont {Sheng}}, \bibinfo {author} {\bibfnamefont
  {Z.}~\bibnamefont {Qu}}, \bibinfo {author} {\bibfnamefont {W.~H.}\
  \bibnamefont {Song}}, \bibinfo {author} {\bibfnamefont {X.~B.}\ \bibnamefont
  {Zhu}},\ and\ \bibinfo {author} {\bibfnamefont {Y.~P.}\ \bibnamefont {Sun}},\
  }\href {https://doi.org/10.1103/PhysRevB.95.245212} {\bibfield  {journal}
  {\bibinfo  {journal} {Phys. Rev. B}\ }\textbf {\bibinfo {volume} {95}},\
  \bibinfo {pages} {245212} (\bibinfo {year} {2017})}\BibitemShut {NoStop}%
\bibitem [{\citenamefont {Liu}\ \emph {et~al.}(2016{\natexlab{a}})\citenamefont
  {Liu}, \citenamefont {Zou}, \citenamefont {Zhang}, \citenamefont {Zhou},
  \citenamefont {Wang}, \citenamefont {Wang}, \citenamefont {Qu},\ and\
  \citenamefont {Zhang}}]{cbCST}%
  \BibitemOpen
  \bibfield  {author} {\bibinfo {author} {\bibfnamefont {B.}~\bibnamefont
  {Liu}}, \bibinfo {author} {\bibfnamefont {Y.}~\bibnamefont {Zou}}, \bibinfo
  {author} {\bibfnamefont {L.}~\bibnamefont {Zhang}}, \bibinfo {author}
  {\bibfnamefont {S.}~\bibnamefont {Zhou}}, \bibinfo {author} {\bibfnamefont
  {Z.}~\bibnamefont {Wang}}, \bibinfo {author} {\bibfnamefont {W.}~\bibnamefont
  {Wang}}, \bibinfo {author} {\bibfnamefont {Z.}~\bibnamefont {Qu}},\ and\
  \bibinfo {author} {\bibfnamefont {Y.}~\bibnamefont {Zhang}},\ }\href@noop {}
  {\bibfield  {journal} {\bibinfo  {journal} {Scientific reports}\ }\textbf
  {\bibinfo {volume} {6}},\ \bibinfo {pages} {1} (\bibinfo {year}
  {2016}{\natexlab{a}})}\BibitemShut {NoStop}%
\bibitem [{\citenamefont {Siberchicot}\ \emph {et~al.}(1996)\citenamefont
  {Siberchicot}, \citenamefont {Jobic}, \citenamefont {Carteaux}, \citenamefont
  {Gressier},\ and\ \citenamefont {Ouvrard}}]{DFTm1}%
  \BibitemOpen
  \bibfield  {author} {\bibinfo {author} {\bibfnamefont {B.}~\bibnamefont
  {Siberchicot}}, \bibinfo {author} {\bibfnamefont {S.}~\bibnamefont {Jobic}},
  \bibinfo {author} {\bibfnamefont {V.}~\bibnamefont {Carteaux}}, \bibinfo
  {author} {\bibfnamefont {P.}~\bibnamefont {Gressier}},\ and\ \bibinfo
  {author} {\bibfnamefont {G.}~\bibnamefont {Ouvrard}},\ }\href@noop {}
  {\bibfield  {journal} {\bibinfo  {journal} {The Journal of Physical
  Chemistry}\ }\textbf {\bibinfo {volume} {100}},\ \bibinfo {pages} {5863}
  (\bibinfo {year} {1996})}\BibitemShut {NoStop}%
\bibitem [{\citenamefont {Motida}\ and\ \citenamefont
  {Miyahara}(1970)}]{supex}%
  \BibitemOpen
  \bibfield  {author} {\bibinfo {author} {\bibfnamefont {K.}~\bibnamefont
  {Motida}}\ and\ \bibinfo {author} {\bibfnamefont {S.}~\bibnamefont
  {Miyahara}},\ }\href {https://doi.org/10.1143/JPSJ.28.1188} {\bibfield
  {journal} {\bibinfo  {journal} {Journal of the Physical Society of Japan}\
  }\textbf {\bibinfo {volume} {28}},\ \bibinfo {pages} {1188} (\bibinfo {year}
  {1970})},\ \Eprint
  {https://arxiv.org/abs/https://doi.org/10.1143/JPSJ.28.1188}
  {https://doi.org/10.1143/JPSJ.28.1188} \BibitemShut {NoStop}%
\bibitem [{\citenamefont {Cheng}\ \emph {et~al.}(2005)\citenamefont {Cheng},
  \citenamefont {Sui}, \citenamefont {Wang}, \citenamefont {Liu}, \citenamefont
  {Miao}, \citenamefont {Huang}, \citenamefont {L{\"u}}, \citenamefont {Qian},\
  and\ \citenamefont {Su}}]{Thi}%
  \BibitemOpen
  \bibfield  {author} {\bibinfo {author} {\bibfnamefont {J.}~\bibnamefont
  {Cheng}}, \bibinfo {author} {\bibfnamefont {Y.}~\bibnamefont {Sui}}, \bibinfo
  {author} {\bibfnamefont {X.}~\bibnamefont {Wang}}, \bibinfo {author}
  {\bibfnamefont {Z.}~\bibnamefont {Liu}}, \bibinfo {author} {\bibfnamefont
  {J.}~\bibnamefont {Miao}}, \bibinfo {author} {\bibfnamefont {X.}~\bibnamefont
  {Huang}}, \bibinfo {author} {\bibfnamefont {Z.}~\bibnamefont {L{\"u}}},
  \bibinfo {author} {\bibfnamefont {Z.}~\bibnamefont {Qian}},\ and\ \bibinfo
  {author} {\bibfnamefont {W.}~\bibnamefont {Su}},\ }\href@noop {} {\bibfield
  {journal} {\bibinfo  {journal} {Journal of Physics: Condensed Matter}\
  }\textbf {\bibinfo {volume} {17}},\ \bibinfo {pages} {5869} (\bibinfo {year}
  {2005})}\BibitemShut {NoStop}%
\bibitem [{\citenamefont {Stoner}\ and\ \citenamefont {Wohlfarth}(1948)}]{Ku}%
  \BibitemOpen
  \bibfield  {author} {\bibinfo {author} {\bibfnamefont {E.~C.}\ \bibnamefont
  {Stoner}}\ and\ \bibinfo {author} {\bibfnamefont {E.}~\bibnamefont
  {Wohlfarth}},\ }\href@noop {} {\bibfield  {journal} {\bibinfo  {journal}
  {Philosophical Transactions of the Royal Society of London. Series A,
  Mathematical and Physical Sciences}\ }\textbf {\bibinfo {volume} {240}},\
  \bibinfo {pages} {599} (\bibinfo {year} {1948})}\BibitemShut {NoStop}%
\bibitem [{\citenamefont {Callen}\ and\ \citenamefont {Callen}(1966)}]{CClaw}%
  \BibitemOpen
  \bibfield  {author} {\bibinfo {author} {\bibfnamefont {H.}~\bibnamefont
  {Callen}}\ and\ \bibinfo {author} {\bibfnamefont {E.}~\bibnamefont
  {Callen}},\ }\href
  {https://doi.org/https://doi.org/10.1016/0022-3697(66)90012-6} {\bibfield
  {journal} {\bibinfo  {journal} {Journal of Physics and Chemistry of Solids}\
  }\textbf {\bibinfo {volume} {27}},\ \bibinfo {pages} {1271} (\bibinfo {year}
  {1966})}\BibitemShut {NoStop}%
\bibitem [{\citenamefont {Khan}\ \emph {et~al.}(2019)\citenamefont {Khan},
  \citenamefont {Zollitsch}, \citenamefont {Arroo}, \citenamefont {Cheng},
  \citenamefont {Verzhbitskiy}, \citenamefont {Sud}, \citenamefont {Feng},
  \citenamefont {Eda},\ and\ \citenamefont {Kurebayashi}}]{FMR1}%
  \BibitemOpen
  \bibfield  {author} {\bibinfo {author} {\bibfnamefont {S.}~\bibnamefont
  {Khan}}, \bibinfo {author} {\bibfnamefont {C.~W.}\ \bibnamefont {Zollitsch}},
  \bibinfo {author} {\bibfnamefont {D.~M.}\ \bibnamefont {Arroo}}, \bibinfo
  {author} {\bibfnamefont {H.}~\bibnamefont {Cheng}}, \bibinfo {author}
  {\bibfnamefont {I.}~\bibnamefont {Verzhbitskiy}}, \bibinfo {author}
  {\bibfnamefont {A.}~\bibnamefont {Sud}}, \bibinfo {author} {\bibfnamefont
  {Y.~P.}\ \bibnamefont {Feng}}, \bibinfo {author} {\bibfnamefont
  {G.}~\bibnamefont {Eda}},\ and\ \bibinfo {author} {\bibfnamefont
  {H.}~\bibnamefont {Kurebayashi}},\ }\href
  {https://doi.org/10.1103/PhysRevB.100.134437} {\bibfield  {journal} {\bibinfo
   {journal} {Phys. Rev. B}\ }\textbf {\bibinfo {volume} {100}},\ \bibinfo
  {pages} {134437} (\bibinfo {year} {2019})}\BibitemShut {NoStop}%
\bibitem [{\citenamefont {Xu}\ \emph {et~al.}(2018)\citenamefont {Xu},
  \citenamefont {Feng}, \citenamefont {Xiang},\ and\ \citenamefont
  {Bellaiche}}]{Kitaev}%
  \BibitemOpen
  \bibfield  {author} {\bibinfo {author} {\bibfnamefont {C.}~\bibnamefont
  {Xu}}, \bibinfo {author} {\bibfnamefont {J.}~\bibnamefont {Feng}}, \bibinfo
  {author} {\bibfnamefont {H.}~\bibnamefont {Xiang}},\ and\ \bibinfo {author}
  {\bibfnamefont {L.}~\bibnamefont {Bellaiche}},\ }\href@noop {} {\bibfield
  {journal} {\bibinfo  {journal} {npj Computational Materials}\ }\textbf
  {\bibinfo {volume} {4}},\ \bibinfo {pages} {1} (\bibinfo {year}
  {2018})}\BibitemShut {NoStop}%
\bibitem [{\citenamefont {Stanley}(1971)}]{CE}%
  \BibitemOpen
  \bibfield  {author} {\bibinfo {author} {\bibfnamefont {H.}~\bibnamefont
  {Stanley}},\ }\href@noop {} {\bibinfo {title} {Introduction to phase
  transitions and critical phenomena oxford univ. press}} (\bibinfo {year}
  {1971})\BibitemShut {NoStop}%
\bibitem [{\citenamefont {Arrott}\ and\ \citenamefont {Noakes}(1967)}]{AP}%
  \BibitemOpen
  \bibfield  {author} {\bibinfo {author} {\bibfnamefont {A.}~\bibnamefont
  {Arrott}}\ and\ \bibinfo {author} {\bibfnamefont {J.~E.}\ \bibnamefont
  {Noakes}},\ }\href@noop {} {\bibfield  {journal} {\bibinfo  {journal}
  {Physical Review Letters}\ }\textbf {\bibinfo {volume} {19}},\ \bibinfo
  {pages} {786} (\bibinfo {year} {1967})}\BibitemShut {NoStop}%
\bibitem [{\citenamefont {Pramanik}\ and\ \citenamefont
  {Banerjee}(2009)}]{AP2}%
  \BibitemOpen
  \bibfield  {author} {\bibinfo {author} {\bibfnamefont {A.~K.}\ \bibnamefont
  {Pramanik}}\ and\ \bibinfo {author} {\bibfnamefont {A.}~\bibnamefont
  {Banerjee}},\ }\href {https://doi.org/10.1103/PhysRevB.79.214426} {\bibfield
  {journal} {\bibinfo  {journal} {Phys. Rev. B}\ }\textbf {\bibinfo {volume}
  {79}},\ \bibinfo {pages} {214426} (\bibinfo {year} {2009})}\BibitemShut
  {NoStop}%
\bibitem [{\citenamefont {Liu}\ \emph {et~al.}(2016{\natexlab{b}})\citenamefont
  {Liu}, \citenamefont {Zou}, \citenamefont {Zhang}, \citenamefont {Zhou},
  \citenamefont {Wang}, \citenamefont {Wang}, \citenamefont {Qu},\ and\
  \citenamefont {Zhang}}]{APCST}%
  \BibitemOpen
  \bibfield  {author} {\bibinfo {author} {\bibfnamefont {B.}~\bibnamefont
  {Liu}}, \bibinfo {author} {\bibfnamefont {Y.}~\bibnamefont {Zou}}, \bibinfo
  {author} {\bibfnamefont {L.}~\bibnamefont {Zhang}}, \bibinfo {author}
  {\bibfnamefont {S.}~\bibnamefont {Zhou}}, \bibinfo {author} {\bibfnamefont
  {Z.}~\bibnamefont {Wang}}, \bibinfo {author} {\bibfnamefont {W.}~\bibnamefont
  {Wang}}, \bibinfo {author} {\bibfnamefont {Z.}~\bibnamefont {Qu}},\ and\
  \bibinfo {author} {\bibfnamefont {Y.}~\bibnamefont {Zhang}},\ }\href@noop {}
  {\bibfield  {journal} {\bibinfo  {journal} {Scientific reports}\ }\textbf
  {\bibinfo {volume} {6}},\ \bibinfo {pages} {1} (\bibinfo {year}
  {2016}{\natexlab{b}})}\BibitemShut {NoStop}%
\bibitem [{\citenamefont {Kouvel}\ and\ \citenamefont {Fisher}(1964)}]{KF}%
  \BibitemOpen
  \bibfield  {author} {\bibinfo {author} {\bibfnamefont {J.~S.}\ \bibnamefont
  {Kouvel}}\ and\ \bibinfo {author} {\bibfnamefont {M.~E.}\ \bibnamefont
  {Fisher}},\ }\href {https://doi.org/10.1103/PhysRev.136.A1626} {\bibfield
  {journal} {\bibinfo  {journal} {Phys. Rev.}\ }\textbf {\bibinfo {volume}
  {136}},\ \bibinfo {pages} {A1626} (\bibinfo {year} {1964})}\BibitemShut
  {NoStop}%
\bibitem [{\citenamefont {Fisher}\ \emph {et~al.}(1972)\citenamefont {Fisher},
  \citenamefont {Ma},\ and\ \citenamefont {Nickel}}]{decay}%
  \BibitemOpen
  \bibfield  {author} {\bibinfo {author} {\bibfnamefont {M.~E.}\ \bibnamefont
  {Fisher}}, \bibinfo {author} {\bibfnamefont {S.-k.}\ \bibnamefont {Ma}},\
  and\ \bibinfo {author} {\bibfnamefont {B.~G.}\ \bibnamefont {Nickel}},\
  }\href {https://doi.org/10.1103/PhysRevLett.29.917} {\bibfield  {journal}
  {\bibinfo  {journal} {Phys. Rev. Lett.}\ }\textbf {\bibinfo {volume} {29}},\
  \bibinfo {pages} {917} (\bibinfo {year} {1972})}\BibitemShut {NoStop}%
\bibitem [{\citenamefont {Zappoli}\ \emph {et~al.}(2015)\citenamefont
  {Zappoli}, \citenamefont {Beysens},\ and\ \citenamefont
  {Garrabos}}]{Ginzburg}%
  \BibitemOpen
  \bibfield  {author} {\bibinfo {author} {\bibfnamefont {B.}~\bibnamefont
  {Zappoli}}, \bibinfo {author} {\bibfnamefont {D.}~\bibnamefont {Beysens}},\
  and\ \bibinfo {author} {\bibfnamefont {Y.}~\bibnamefont {Garrabos}},\
  }\bibinfo {title} {The ginzburg criterion},\ in\ \href
  {https://doi.org/10.1007/978-94-017-9187-8_19} {\emph {\bibinfo {booktitle}
  {Heat Transfers and Related Effects in Supercritical Fluids}}}\ (\bibinfo
  {publisher} {Springer Netherlands},\ \bibinfo {address} {Dordrecht},\
  \bibinfo {year} {2015})\ pp.\ \bibinfo {pages} {371--372}\BibitemShut
  {NoStop}%
\bibitem [{\citenamefont {Cai}\ \emph {et~al.}(2020)\citenamefont {Cai},
  \citenamefont {Sun}, \citenamefont {Xia}, \citenamefont {Wu}, \citenamefont
  {Liu}, \citenamefont {Liu}, \citenamefont {Gong}, \citenamefont {Yao},
  \citenamefont {Guo},\ and\ \citenamefont {Wang}}]{SC}%
  \BibitemOpen
  \bibfield  {author} {\bibinfo {author} {\bibfnamefont {W.}~\bibnamefont
  {Cai}}, \bibinfo {author} {\bibfnamefont {H.}~\bibnamefont {Sun}}, \bibinfo
  {author} {\bibfnamefont {W.}~\bibnamefont {Xia}}, \bibinfo {author}
  {\bibfnamefont {C.}~\bibnamefont {Wu}}, \bibinfo {author} {\bibfnamefont
  {Y.}~\bibnamefont {Liu}}, \bibinfo {author} {\bibfnamefont {H.}~\bibnamefont
  {Liu}}, \bibinfo {author} {\bibfnamefont {Y.}~\bibnamefont {Gong}}, \bibinfo
  {author} {\bibfnamefont {D.-X.}\ \bibnamefont {Yao}}, \bibinfo {author}
  {\bibfnamefont {Y.}~\bibnamefont {Guo}},\ and\ \bibinfo {author}
  {\bibfnamefont {M.}~\bibnamefont {Wang}},\ }\href
  {https://doi.org/10.1103/PhysRevB.102.144525} {\bibfield  {journal} {\bibinfo
   {journal} {Phys. Rev. B}\ }\textbf {\bibinfo {volume} {102}},\ \bibinfo
  {pages} {144525} (\bibinfo {year} {2020})}\BibitemShut {NoStop}%
\bibitem [{\citenamefont {Xu}\ \emph {et~al.}(2020)\citenamefont {Xu},
  \citenamefont {Feng}, \citenamefont {Kawamura}, \citenamefont {Yamaji},
  \citenamefont {Nahas}, \citenamefont {Prokhorenko}, \citenamefont {Qi},
  \citenamefont {Xiang},\ and\ \citenamefont {Bellaiche}}]{QSL}%
  \BibitemOpen
  \bibfield  {author} {\bibinfo {author} {\bibfnamefont {C.}~\bibnamefont
  {Xu}}, \bibinfo {author} {\bibfnamefont {J.}~\bibnamefont {Feng}}, \bibinfo
  {author} {\bibfnamefont {M.}~\bibnamefont {Kawamura}}, \bibinfo {author}
  {\bibfnamefont {Y.}~\bibnamefont {Yamaji}}, \bibinfo {author} {\bibfnamefont
  {Y.}~\bibnamefont {Nahas}}, \bibinfo {author} {\bibfnamefont
  {S.}~\bibnamefont {Prokhorenko}}, \bibinfo {author} {\bibfnamefont
  {Y.}~\bibnamefont {Qi}}, \bibinfo {author} {\bibfnamefont {H.}~\bibnamefont
  {Xiang}},\ and\ \bibinfo {author} {\bibfnamefont {L.}~\bibnamefont
  {Bellaiche}},\ }\href {https://doi.org/10.1103/PhysRevLett.124.087205}
  {\bibfield  {journal} {\bibinfo  {journal} {Phys. Rev. Lett.}\ }\textbf
  {\bibinfo {volume} {124}},\ \bibinfo {pages} {087205} (\bibinfo {year}
  {2020})}\BibitemShut {NoStop}%
\end{thebibliography}
\providecommand{\noopsort}[1]{}\providecommand{\singleletter}[1]{#1}%

\end{document}